\documentclass[preprint,aps,prd,
nofootinbib,superscriptaddress,
tightenlines]{revtex4}
\usepackage{bm}

\usepackage{amsmath}
\usepackage{float}
\usepackage{dcolumn}
\usepackage{amssymb}
\usepackage{graphics}
\usepackage{amsmath, amsfonts, amsthm, amssymb, graphicx,comment}
\usepackage{etex}
\usepackage{hyperref,subfigure}
\usepackage{footnote}
\usepackage{xcolor}

\newcommand{\bea}{\begin{eqnarray}}
\newcommand{\eea}{\end{eqnarray}}

\allowdisplaybreaks[3]

\def \bal#1\eal  {\begin{align} #1 \end{align}}
\newcommand{\be} {\begin{equation}}
\newcommand{\ee} {\end{equation}}
\newcommand{\bpm}{\begin{pmatrix}}
\newcommand{\epm}{\end{pmatrix}}
\newcommand{\nn} {\nonumber\\}

\newcommand{\ud} {\mathrm{d}}

\newcommand{\pd} {\partial}

\newcommand{\mc} {\mathcal}

\newcommand{\ai}{{\alpha}}
\newcommand{\bi}{{\beta}}

\newcommand{\ri}{{\rho}}
\newcommand{\si}{{\sigma}}

\begin{document}

\title{No hair theorem in quasi-dilaton massive gravity}

\author{De-Jun Wu}
\email[]{wudejun10@mails.ucas.ac.cn}
\affiliation{School of Physics, University of Chinese Academy of Sciences, Beijing 100049, China}

\author{Shuang-Yong Zhou}
\email[]{sxz353@case.edu}
\affiliation{Department of Physics, Case Western Reserve University, 10900 Euclid Ave, Cleveland, OH 44106, USA}

\begin{abstract}

We investigate the static, spherically symmetric black hole solutions in the quasi-dilaton model and its generalizations, which are scalar extended dRGT massive gravity with a shift symmetry. We show that, unlike generic scalar extended massive gravity models, these theories do not admit static, spherically symmetric black hole solutions until the theory parameters in the dRGT potential is fine-tuned. When fine-tuned, the geometry of the static, spherically symmetric black hole is necessarily that of general relativity and the quasi-dilaton field is constant across the spacetime. The fine-tuning and the no hair theorem apply to black holes with flat, anti-de Sitter or de Sitter asymptotics.

\end{abstract}

\maketitle
\section{Introduction}

Generic massive gravity theories are known to suffer from theoretical pathologies such as the Boulware-Deser ghost  \cite{Boulware:1973my}, van Dam-Veltman-Zakharov (vDVZ) discontinuity \cite{vanDam:1970vg,Zakharov:1970cc} and low strong coupling scale ($\Lambda_5=(M_Pm^2)^{1/5}$) \cite{ArkaniHamed:2002sp}. However, a unique two-parameter family massive gravity theory, the de Rham-Gabadadze-Tolley (dRGT) model, has been found \cite{deRham:2010ik, deRham:2010kj} that is free of the Boulware-Deser ghost \cite{deRham:2010ik, deRham:2010kj, Hassan:2011hr, Hassan:2011ea} (see \cite{deRham:2014zqa,Hinterbichler:2011tt} for recent reviews). The dRGT model has a $\Lambda_3=(M_Pm^2)^{1/3}$ strong coupling scale around the Minkowski vacuum, on which the vDVZ discontinuity is persistent, but a consistent Vainshtein mechanism has been argued to recover general relativity within most conventional circumstances that have been experimentally tested \cite{deRham:2014zqa,Hinterbichler:2011tt}. More recently, it has been realized that the dRGT model actually has near-flat backgrounds that are totally free of the vDVZ discontinuity and where the strong coupling scale can be raised to a quasi-$\Lambda_2=(M_Pm)^{1/2}$ scale \cite{deRham:2015ijs,deRham:2016plk}.

Various aspects of the black holes in the dRGT model and its extensions have been investigated in the literature \cite{Koyama:2011xz,Gruzinov:2011mm, Deffayet:2011rh,Comelli:2011wq,Berezhiani:2011mt,Volkov:2012wp,Baccetti:2012ge,Cai:2012db,Berezhiani:2013dw,Mirbabayi:2013sva,Volkov:2013roa,Babichev:2013una,Brito:2013wya,Berezhiani:2013dca,Brito:2013xaa,Kodama:2013rea,Babichev:2014oua,Babichev:2014fka,Babichev:2014tfa,Volkov:2014ooa,Babichev:2015xha,Kobayashi:2015yda,Enander:2015, Tolley:2015ywa}. The simplest black holes are those with spherical symmetry and time translational symmetry. It has been shown that the static and spherically symmetric black holes in the original dRGT model are exactly of the same form of those in general relativity (GR-like) \cite{Volkov:2012wp,Volkov:2014ooa}. That is, making the graviton massive (adding the helicity-1 and helicity-0 polarizations to the graviton) does not endow extra hair to the static, spherically symmetric black holes.

Assuming flat asymptotics, the Schwarzschild geometry is the unique static, spherically symmetric black hole in general relativity, the only parameter being the mass of the black hole. Relaxing the staticity and the spherical symmetry can add the angular momentum as a new parameter, and adding the Maxwell field can add another -- the electromagnetic charge. However, it is typical of a gravitational theory that adding extra fields usually does not induce extra features for the black hole solutions, thanks to a range of no hair theorems (see, e.g.,~\cite{Bekenstein:1995un, Sotiriou:2011dz,Hui:2012qt}). But it is also known that there are cases where this is violated. For example, for the case of a general second order scalar-tensor theory with a shift symmetry, a black hole will necessarily be hairy if the Lagrangian contains a linear coupling between the scalar and the Gauss-Bonnet invariant  \cite{Sotiriou:2013qea, Sotiriou:2014pfa}. See \cite{Bekenstein:1996pn,Herdeiro:2015waa} for a review of no hair theorems and the exceptional hairy cases.

However, in the context of dRGT massive gravity, the GR-like black hole have been found to suffer from undesirable features such as physical singularities on the horizon, instabilities or strong couplings \cite{Berezhiani:2011mt,Berezhiani:2013dw,Berezhiani:2013dca,Kodama:2013rea}. In extended dRGT models, the black hole solutions are richer and hopefully better behaved. A simple way to extend the dRGT model is to add an extra scalar field \cite{Huang:2012pe, D'Amico:2011jj,Huang:2013mha,D'Amico:2012zv,DeFelice:2013dua,DeFelice:2013tsa}. For example, one may promote the graviton mass to depend on the extra scalar as in mass-varying massive gravity \cite{Huang:2012pe, D'Amico:2011jj}, or one may add a scalar field with a new global symmetry as in quasi-dilaton massive gravity and its generalizations \cite{D'Amico:2012zv,DeFelice:2013dua,DeFelice:2013tsa}. In mass-varying massive gravity, hairy black holes with flat or anti-de Sitter asymptotics have been found \cite{Tolley:2015ywa}. Unlike the bi-gravity extension of the dRGT model \cite{Babichev:2015xha}, an interesting feature in mass-varying massive gravity is that generically there is no need to fine-tune the theory parameters to get black holes with flat asymptotics, as long as the effective scalar potential has a local minimum \cite{Tolley:2015ywa}.

The extra scalar field in quasi-dilaton massive gravity enjoys a combined global (shift) symmetry:
\be
\label{gsym}
\sigma\rightarrow\sigma+\sigma_0,~~~~~~
\phi^a\rightarrow e^{-\sigma_0}\phi^a.
\ee
where $\sigma$ is the quasi-dilaton, $\phi^a$ are the Stueckelberg fields of the massive graviton and $\sigma_0$ is a constant. The original quasi-dilaton massive gravity does not admit stable homogenous and isotropic solutions \cite{Gumrukcuoglu:2013nza}, much like the simplest dRGT model. The generalized quasi-dilaton model was subsequently proposed to render the homogenous and isotropic background stable \cite{DeFelice:2013tsa}. However, we stress that, also like the dRGT model \cite{deRham:2015ijs,deRham:2016plk}, quasi-dilaton massive gravity may still have phenomenologically acceptable inhomogeneous and/or anisotropic cosmological solutions.

In this paper, we investigate the static and spherically symmetric black holes in the quasi-dilaton and generalized quasi-dilaton massive gravity. By examining the equations of motion at infinity, we will see that these theories do not admit a regular black hole solution until one of the theory parameter is fine-tuned. This is closely related to the global symmetry of the model. More specifically, as we shall see that it is essentially because the quasi-dilaton models have a run-away ``effective'' scalar potential, unless the dRGT potential parameters are fine-tuned. When fine-tuned, the effective potential cancels out. By examining the equations of motion on the black hole event horizon, we will see that the quasi-dilaton is necessarily a constant field and the metric then takes one of the forms of the corresponding black holes in general relativity. (Note that while in general relativity a coordinate transform of a solution is just the same solution in a different gauge, the coordinated transformed solution in massive gravity (in unitary gauge) is a physically different solution.) Therefore, the quasi-dilaton field does not endow any extra hair for the static, spherically symmetric black holes. These results hold for black holes with flat, anti-de Sitter or de Sitter asymptotics.  In Section \ref{sec:qdmg}, we will discuss the fine-tuning and no hair theorem in the original quasi-dilaton model. The same steps can be taken to show the fine-tuning and no hair theorem in the generalized quasi-dilaton model, except that the equations of motion are more involved, which will be shown succinctly in Section \ref{sec:gqdmg}.

\section{Quasi-dilaton massvie gravity}
\label{sec:qdmg}

We will start with quasi-dilation massive gravity \cite{D'Amico:2012zv} whose equations are relatively simple to work with. In Section \ref{sec:gqdmg}, we will extend our discussion to generalized quasi-dilaton massive gravity. The essential steps of arguments are essentially the same for the two models, the only difference being that the equations of motion in the later case is more involved.

Quasi-dilaton massive gravity is given by \cite{D'Amico:2012zv}
\be
\label{theth}
S = \frac{{{M_{P}^2} }}{2} \!\int \!\!  {d^4 x} \sqrt { - g} \left[ {R + m^2 U({\mathcal{K}}) - \frac{\omega}{2}  \partial _\mu  \sigma         \partial^\mu   \sigma}        \right],
\ee
where $U(\mc{K})\equiv U_2  + \alpha _3 U_3  + \alpha _4 U_4$ and
\be
U_2  = {\cal K}_{[\mu }^\mu  {\cal K}_{\nu ]}^\nu  , ~~U_3  = {\cal K}_{[\mu }^\mu  {\cal K}_\nu ^\nu  {\cal K}_{\rho ]}^\rho  , ~~U_4  = {\cal K}_{[\mu }^\mu  {\cal K}_\nu ^\nu  {\cal K}_\rho ^\rho  {\cal K}_{\sigma ]}^\sigma   ,
\ee
with
\be
\label{defk}
{\cal K}_\nu ^\mu   = \delta _\nu ^\mu   - e^{\sigma}\sqrt {g^{\mu \rho }  \pd_\ri\phi^a\pd_\nu\phi^b\eta_{ab}},
\ee
$\eta_{\ri\nu}$ is the reference Minkowski metric, $m$ is the graviton mass parameter, $\alpha_3$ and $\alpha_4$ are two free parameters, and $\omega$ is a positive constant. $\phi^a$ are the Stueckelberg fields introduced to restore the diffeomorphism invariance. As can be easily seen, there is a global symmetry for the combined operation on $\sigma$ and $\phi^a$: $\sigma\rightarrow\sigma+\sigma_0,~\phi^a\rightarrow e^{-\sigma_0}\phi^a$. As it is easier to work with for explicit solutions, we shall adopt the unitary gauge in the rest of the paper
\be
\phi^a=x^a .
\ee
The vacuum Einstein field equation is given by
\bal
\label{eisteq}
G_{\mu\nu} =\frac{1}{2}T^{(\sigma)}_{\mu\nu}+m^2 X_{\mu\nu} ,
\eal
where
\be
T^{(\sigma)}_{\mu\nu} =\omega\left( \partial_\mu \sigma \partial _\nu \sigma  - g_{\mu\nu} \frac{1}{2}\partial _\ri  \sigma \partial^\ri   \sigma\right),
\ee
and
\bal
 X_{\mu\nu}  &= \frac{1}{2}\bigg[-({\cal K}_{\mu\nu}  - [{\cal K}] g_{\mu\nu})  + \alpha \left( {{\cal K}_{\mu\nu}^2  - [{\cal K}] {\cal K}_{\mu\nu}  + U_2 g_{\mu\nu} } \right)
 \nn
&~~~~~~  - \beta \left( {{\cal K}_{\mu\nu}^3  - [{\cal K}] {\cal K}_{\mu\nu}^2  + U_2 {\cal K}_{\mu\nu}  - U_3 g_{\mu\nu} } \right)\bigg]  .
\eal
Note that we have introduced two new parameters to replace $\alpha_3$ and $\alpha_4$:
\be
\label{abequ}
\alpha=1+\alpha_3, ~~~~ \beta=\alpha_3+\alpha_4  ,
\ee
which we will use in the rest of the paper. The $\sigma$ equation of motion can be written as
\be
\label{eompsi}
\partial _\mu \left( {\sqrt {-g } g^{\mu\nu}  \partial_\nu \sigma } \right) + m^2 {\sqrt { -g } }\partial _\sigma U=0.
\ee
For our later convenience, we can also re-write $U_2$, $U_3$ and $U_4$ in terms of the four eigenvalues of $\cal K_\nu^\mu$ (labeled as $k_1, k_2, k_3$ and $k_4$):
\be
\label{defU2}
U_2={\sum\limits_{\nu<\mu} {{k_\nu}{k_\mu}} },~
U_3=\!\!\!{\sum\limits_{\nu<\mu<\rho}\!\! {{k_\nu}{k_\mu}{k_\rho}}},~
U_4={k_1}{k_2}{k_3}{k_4}  .
\ee

\subsection{Staticity and spherical symmetry}

We are interested in static, spherically symmetric black holes.  As mentioned earlier, we will work in unitary gauge, so the general static, spherically symmetric ansatz is given by
\begin{align}
\label{metric}
\ud s^2  &=  - a(r)dt^2 + 2b(r)drdt  + e(r)dr^2   + f(r)d\Omega ^2,
\\
\label{metric2}
\ud s^2_\eta &= \eta_{\mu\nu}dx^{\mu}dx^{\nu}  =  - dt^2  + dr^2   +r^2 d\Omega ^2  ,
\\
\label{varphian}
\sigma &= \sigma(r)    .
\end{align}

The ${}^r_t$ component of the Einstein equation is simply
\begin{align}
\label{fenlei}
{b\left( {\beta {{k_3}}^2  + 2\alpha {{k_3}} + 1} \right)=0}  ,
\end{align}
which gives two branches of solutions: (i) $b=0$; (ii) $\beta {{k_3}}^2  + 2\alpha {{k_3}} + 1 = 0$. The (i) branch leads to black holes with physical singularities on the event horizon \cite{Berezhiani:2011mt,Deffayet:2011rh}. So it is necessary to have $b\neq0$ to avoid this.  Therefore, we shall only consider the (ii) branch, which fixes $k_3$ to be constant
\be
\label{k3eq}
{{k_3}} = \frac{\pm\sqrt{\alpha ^2-\beta }-\alpha }{\beta }  .
\ee
The the reality of $k_3$ additionally requires: $\ai^2>\bi$. Together with the definition of $k_3$, this also fixes $f(r)$ to be
\be
f(r)= \frac{ e^{2 \sigma }}{\left(1-k_3\right){}^2}r^2 ,
\ee
and also significantly simplifies the rest components of the Einstein equation. The ${}^t_r$, ${}^t_t$ and ${}^\theta_\theta$ components are respectively given by
\bea
\label{eomc}
2( r \sigma '+1)c'&=&\left(4 r \sigma ''+ \left( (\omega +4)r \sigma '+8\right)\sigma '\right) c,\\
\label{eoma}
4 r \left(r \sigma '+1\right)a'&=&  \left(  \left( (\omega -4)r \sigma '-8\right)r\sigma '-4\right)a
\nn
&~&+2 k_3 r^2  m^2 \left(\alpha  k_3+2\right)c
\nn
&~&+4 \left(k_3-1\right){}^2 ce^{-2 \sigma }.
\\
\label{eom3}
2rc a'' &=& - \left( {4c\left( {r\sigma ' + 1} \right) - rc'} \right)a' \nn
&~&+ 2m^2 rc^2 \big(   \alpha (k_1 k_2  + k_1 k_3  + k_2 k_3 )\nn
&~& + k_1  + k_2  + k_3 +\beta k_1 k_2 k_3 \big)
\eea
where $c\equiv ae+b^2$. By spherical symmetry, we also have $k_4=k_3$. $k_1$ and $k_2$ are further related to the unknowns by $\det(I-{\mc{K}})=(1-k_1)(1-k_2)(1-k_3)(1-k_4)$, which gives
\be
\label{k1k2}
k_1 k_2 -( k_1+k_2 )=\frac{e^ {2 \sigma }}{\sqrt{c}}-1.
\ee
We can also derive the $\sigma$ equation of motion, by making use of the relation $\partial k_n/\partial \sigma=k_n-1$. Combining it with Eq.~\eqref{eomc}, \eqref{eoma}, \eqref{eom3} and \eqref{k1k2}, we can arrive at
\be
\label{eomj}
\sigma''-B r^2 \sigma'^2+  (A -2B r) \sigma' =  B,
\ee
where
\bea
A&=&\!\frac{ \left( m^2 r^2 \left(\alpha  k_3+2\right)k_3 \!+\!2 \left(k_3-1\right)^2\! e^{-2 \sigma }\right)\!c+2 a}{2ra} ,
\\
B&=&\frac{4  m^2 \left((\alpha -1) k_3+2\right) k_3   e^{ 2\sigma } \sqrt{c}}{ \omega a}.
\eea
$B$ may be roughly interpreted as an ``effective'' potential for the scalar field $\sigma$, and we see that this effective potential is of the run-away type (of the type $\sim e^{2\sigma}$), if the constant in it does not vanish identically. Eq.~(\ref{eomj}) will be the key to understand the parameter fine-tuning and no hair theorem in the following.

\subsection{Fine-tuning of theory parameters}

In this section we will study the perturbative solutions near the spatial infinity and near the black hole event horizon. This allows us to show that one needs to fine-tune a theory parameter to get a static, spherically symmetric black hole solution ($\ai$ and $\bi$ satisfying Eq.~(\ref{abeqty})), and when such a black hole exists it does not have any scalar hair from $\sigma$.

Let us start with the perturbative solutions near the spatial infinity. First, note that at the spatial infinity we have $\sigma$ tending to a constant, otherwise the gradient energy of $\sigma$ would diverge, making the whole solution unphysical\,\footnote{In particular, the case of $\sigma(r \to +\infty)\to-\infty$ would lead to an infinite gradient energy from the Lagrangian term $-\sqrt{-g}\pd_\mu\sigma \pd^\mu \sigma$. This is because, to have $\sigma(r \to +\infty)\to-\infty$, we would need $\sigma\sim -r^\eta$ with constant $\eta>0$, which gives rise to an infinite gradient energy $E^{\si}_{\rm gradient} \sim  \int\! \ud r r^2 [\sigma'(r)]^2 \sim \int\! \ud r r^{2\eta}\to +\infty$. To have a finite $E^{\si}_{\rm gradient}$, it is necessary that $\eta\leq 0$, which corresponds to $\sigma(r \to +\infty)\to {\rm const}$. (The finite gradient energy requirement actually imposes stricter condition on $\eta$ or the next leading order of $\sigma(r)$. If further assuming $\si(r)$ to be analytic at spatial infinity, we have the expansion of Eq.~(\ref{acsinf}).)}. By Eq.~\eqref{eomj}, we then see that for a black hole solution to exist we need
\be
\label{Bcon}
B \to 0, ~~~{\rm as}~~r\to \infty .
\ee
For black holes with flat asymptotics, this directly implies
\be
\label{jcond}
\left((\alpha -1) k_3+2\right) k_3=0 ,
\ee
since in this case $m^2$, $\omega$, $e^{2\sigma}$ and $\sqrt{c}$ \footnote{If $c=0$, the metric becomes singular.} are strictly nonzero and $a$ is finite at the spatial infinity. That is, the theory parameters must be fine-tuned to cancel out the effective potential.

Eq.~(\ref{jcond}) also holds for black holes with anti-de Sitter or de Sitter asymptotics, but it needs slightly more algebra to see. The complication comes from the fact that $a$ goes like $r^2$ at infinity for these asymptotics. So $B$ could go to zero at the spatial infinity simply because of this behavior of $a$. Thus, for the anti-de Sitter or de Sitter case, one shall examine the perturbative solution near the spatial infinity more carefully. To this end, we expand the unknown functions around the spatial infinity as:
\bal
\label{acsinf}
a({r})&=a_{-2} r^2+a_0+\frac{a_1}{r}+\frac{a_2}{r^2}+\cdots,
\nn
c({r})&=c_0+\frac{c_1}{r}+\frac{c_2}{r^2}+\cdots,
\nn
\sigma ({r})&=\sigma _0+\frac{\sigma _1}{r}+\frac{\sigma _2}{r^2}+\cdots .
\eal
Solving the equations of motion perturbatively at the spatial infinity, we find that the leading order of  Eq.~(\ref{eomj}) is given by
\be
\label{loeomj}
\frac{4  m^2 \left((\alpha -1) k_3+2\right) k_3   e^{2 \sigma_0 } }{ \sqrt{c_0} } =0 .
\ee
This again implies Eq.~(\ref{jcond}). (As expected, Eq.~(\ref{loeomj}) also holds for flat asymptotics.)

\begin{figure}[!htb]\centering
\includegraphics[width=.3\textwidth]{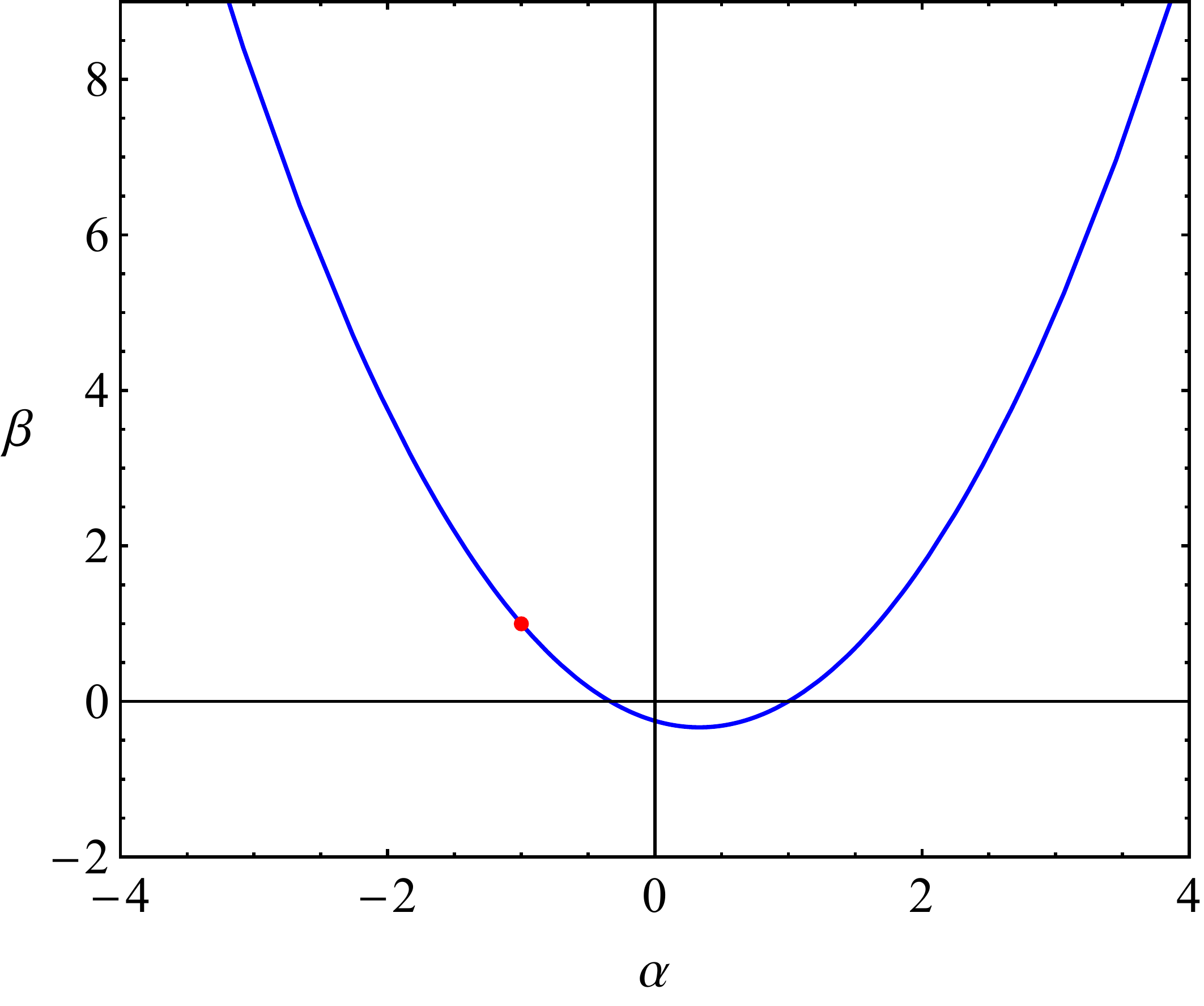}
\caption{Fine-tuning of the theory parameters on the $\ai$-$\bi$ plane. For a static, spherically symmetric black hole to exist, the theory parameters have to lie on the blue line: $4\beta = -1 - 2 \alpha + 3 \alpha^2$. The red dot at $(-1,1)$ has to be marked off as it implies a divergent $f(r)$. This plot is the same for the quasi-dilaton model and the generalized quasi-dilaton model. Note that $\alpha=1+\alpha_3, \beta=\alpha_3+\alpha_4$.}
\label{p1}
\end{figure}

Therefore, Eq.~(\ref{jcond}) holds regardless of the asymptotics of the black hole. Together with Eq.~\eqref{k3eq}, this implies
\be
\label{k3neq0}
k_3=\frac{2}{1-\ai}\neq 0,
\ee
and gives rise to an equality between $\ai$ and $\bi$. The finiteness of $f(r)$ requires that $k_3\neq 1$, meaning that $\ai\neq -1$. Thus, our first result is that, for a static, spherically symmetric black hole to exist, one needs to fine-tune the theory parameters such that
\be
\label{abeqty}
4\beta = -1 - 2 \alpha + 3 \alpha^2 , ~(  \ai \neq-1).
\ee
where $\ai$ and $\bi$ are given by Eq.~(\ref{abequ}). See Fig.~(\ref{p1}).

This is reminiscent of black hole solutions in massive bi-gravity \cite{Babichev:2015xha}, where a fine-tuning between the two theory parameters are required to have an asymptotically flat black hole. However,  in a generic scalar extended dRGT massive gravity, this kind of fine-tuning is not required, as long as the effective scalar potential has a local minimum \cite{Tolley:2015ywa}. The peculiarity of quasi-dilaton massive gravity is that it has a run-away effective potential if not fine-tuned. When the theory parameters are fine-tuned, as we shall assume in the following, the effective potential cancels out, which leads to a no hair theorem of the black holes.

\subsection{No hair theorem}

Now, we turn to the perturbative solution on the black hole event horizon $r=r_h$. Since a black hole is assumed to exist (i.e., Eq.~\eqref{jcond} is imposed), Eq.~(\ref{eomj}) reduces to
\be
\label{eomjj}
\sigma''= - A  \sigma'   .
\ee
For the black hole ansatz (\ref{metric}), it is a geometric fact that the event horizon should be a Killing horizon of vector $\pd_t$, so we have $a(r_h)=-(g_{\mu\nu} (\pd_t)^\mu(\pd_t)^\nu )_{r_h}=0$. Since $A\propto 1/a$, $A  \sigma'$ could diverge on the horizon, which by Eq.~(\ref{eomjj}) would imply that $\sigma''$ and hence $\sigma$ itself diverges on the horizon. Since $\sigma$ is a scalar, this amounts to a physical singularity. To avoid this, we need to impose that at the horizon $r=r_h$ we have
\be
\left.\left(  m^2 r^2 \left(\alpha  k_3+2\right)k_3+2 \left(k_3-1\right)^2\! e^{-2 \sigma }\right)\sigma'\right|_{r_h}=0.
\ee
Note that we have used the fact that $c\neq0$ anywhere. Now, by Eq.~(\ref{jcond}) and Eq.~(\ref{k3neq0}), we see that $k_3  m^2 r^2 \left(\alpha  k_3+2\right)+2 \left(k_3-1\right)^2\! e^{-2 \sigma }$ is strictly positive. Thus, we have
\be
\sigma'(r_h)=0.
\ee

Then, by Eq.~(\ref{eomjj}), we see that $\sigma''(r_h)=0$. Acting $'$ on Eq.~(\ref{eomjj}), we can express $\sigma'''$ in terms of
\be
\sigma'''= A\sigma''+ A' \sigma'.
\ee
Since $\sigma'(r_h)=\sigma''(r_h)=0$, we have $\sigma'''(r_h)=0$.  Repeatedly acting $'$ on Eq.~(\ref{eomjj}), it is easy to see that
\be
\left.\frac{\ud^n\sigma}{\ud r^n}\right|_{r_h}=0,~~{\rm for~any}~~n .
\ee
Since $\sigma$ is an analytic function of $(r-r_h)$, this means that
\be
\label{src}
\sigma(r)=const.
\ee
Now, Eq.~(\ref{src}) reduces the equation of motion to the case of the dRGT model, and we know that in the dRGT model the only static, spherically symmetric black hole solutions are those of general relativity \cite{Volkov:2012wp,Volkov:2014ooa}. Thus, we have proved that static, spherically symmetric black holes in quasi-dilaton massive gravity can not have the scalar hair of $\sigma$.

\section{Extension to generalized quasi-dilaton massive gravity}
\label{sec:gqdmg}

Generalized quasi-dilaton massive gravity has been shown to have stable homogeneous and isotropic solutions \cite{DeFelice:2013tsa}, making it more appealing for cosmological applications, at least at a practical level. Here we will show the same fine-tuning and no hair theorem apply to this generalization.

The generalized quasi-dilaton model is obtained by the replacement in the quasi-dilaton model (\ref{theth}):
\be
\pd_\mu\phi^a \pd_\nu\phi^b \eta_{ab} \to  \pd_\mu\phi^a \pd_\nu\phi^b \eta_{ab}  - \ai_\si e^{-2\si}\pd_\mu \sigma \pd_\nu \sigma ,
\ee
where $\ai_\si$ is constant; Or, in unitary gauge, by the replacement:
\be
\label{repuni}
\eta_{\mu\nu}  \to \eta_{\mu\nu}  - \ai_\si e^{-2\si}\pd_\mu \sigma \pd_\nu \sigma .
\ee
Note that in generalized quasi-dilaton massive gravity the global symmetry (\ref{gsym}) is kept intact.

Substituting the ansatz (\ref{metric}), (\ref{metric2}) and (\ref{varphian}) into the ${}^r_t$ component of the Einstein equation, we again obtain Eq.~(\ref{fenlei}). Again, since $b\neq 0$, $k_3$ is given by Eq.~(\ref{k3eq}). The forms of the ${}^t_r$, ${}^t_t$ and ${}^\theta_\theta$ components of the Einstein equation also superficially remain unchanged in this generalization, but $k_1$ and $k_2$ are now related to the unknowns via
\be
k_1k_2 -( k_1+k_2 ) = \frac{e ^{2 \sigma } \sqrt{1-\alpha_\sigma  e^ { -2 \sigma}  \sigma '^2}}{\sqrt{c}}-1.
\ee
However, the equivalent of Eq.~(\ref{eomj}) is now more complicated:
\be
\label{Feom}
F(\sigma'',\si',a,c;m,\ai,\bi,\omega,\ai_\si) =0 .
\ee
The cumbersome expression of the function $F$ can be easily obtained with a computer algebra system.

\subsection{Fine-tuning of theory parameters}

To see we will still require the fine-tuning of the theory parameters to have static, spherically symmetric black holes, we substitute the ansatz at the spatial infinity (\ref{acsinf}) into the equations of motion, and find that the leading order of Eq.~(\ref{Feom}) is again Eq.~(\ref{loeomj}). Note that Eq.~(\ref{loeomj}) does not contain $a_{-2}$, so it also applies to flat asymptotics. This again implies the fine-tuning condition (\ref{k3eq}). See Fig.~(\ref{p1}).

\subsection{No hair theorem}

Substituting the fine-tuning condition (\ref{abeqty}) into Eq.~(\ref{Feom}), we get exactly Eq.~(\ref{eomjj}). The rest of the no hair proof of quasi-dilaton massive gravity can then be carried over literally. Therefore, there is also a no hair theorem in generalized quasi-dilaton massive gravity for static, spherically symmetric black holes.

\section{Discussions}

The global symmetry in the quasi-dilaton models significantly constrains the form of the Lagrangian that can be constructed. The requirement of the existence of a static, spherically symmetric black hole further imposes a constraint on the parameter space of the model. That is, to allow for a static, spherically symmetric black hole to exist, the two-parameter theory space ($\ai$ and $\bi$) of dRGT massive gravity is confined to a one dimensional subspace constrained by $4\beta = -1 - 2 \alpha + 3 \alpha^2 ,~(  \ai \neq-1)$.

Furthermore, we have shown that the static, spherically symmetric black holes in the quasi-dilaton models have no hair of the quasi-dilaton field: The quasi-dilaton field must be constant across the spacetime and the metric must take the corresponding form in general relativity. This is regardless of whether the asymptotics are flat, de Sitter or anti-de Sitter.

Quasi-dilaton massive gravity is a special case of generalized mass-varying massive gravity \cite{Huang:2013mha}. Generically, generalized mass-varying massive gravity has hairy black hole solutions \cite{Tolley:2015ywa}. Besides quasi-dilaton massive gravity, it may be expected that there are some other special cases where black holes do not bear extra scalar hair. One may wonder whether our proof of the no hair theorem can be used to find some of those other special cases. In generalized mass-varying massive gravity, the corresponding scalar equation of motion goes like $\sigma''+C\sigma'^2+D\sigma'+E=0$. For a black hole solution to exist, one requires that $E$ have a zero at the spatial infinity. If it happens that $E$ can not reach zero, i.e., the scalar effective potential does not have a local minimum, then some theory parameters may have to be fine-tuned to accommodate the black hole solution. If in addition $C$ vanishes either by the theory space fine-tuning or by construction, then we can similarly proceed to prove a no hair theorem in the corresponding model. Of course, it does not mean the black hole solution definitely has scalar hair for a model that can not be shown to have a no hair theorem with this method.

Generalized quasi-dilaton massive gravity has been further generalized \cite{DeFelice:2013dua} to include shift-symmetric Horndeski terms  as well as a special quasi-dilatation term $e^{4\si} \sqrt{-\det(\eta_{\mu\nu}  - \ai_\si e^{-2\si}\pd_\mu \sigma \pd_\nu \sigma)}$. Let us consider the theory without this special quasi-dilatation term first. Expanding around the spatial infinity,  all the shift-symmetric Horndeski terms do not contribute to Eq.~(\ref{loeomj}). (One potential term that may contribute is the linear coupling between $\sigma$ and the Gauss-Bonnet invariant \cite{Sotiriou:2013qea}; But this term only contributes to higher orders in $1/r$, because the Gauss-Bonnet invariant vanishes for a constant metric.) Therefore, there will still be the fine-tuning between $\ai$ and $\bi$ for a static, spherically symmetric black hole to exist. Although a no hair theorem can be established \cite{Hui:2012qt} in shift-symmetric Horndeski theory that excludes the linear coupling between $\sigma$ and the Gauss-Bonnet invariant  \cite{Sotiriou:2013qea}, adding shift-symmetric Horndeski terms to generalized quasi-dilation massive gravity \cite{DeFelice:2013tsa} breaks the proof chain above. The reason is that there will be some $(\si')^n$ terms added to Eq.~(\ref{eomjj}), which will lead to solutions other than $\sigma'=0$. Finally, once the special quasi-dilatation term is added to the Lagrangian, Eq.~(\ref{loeomj}) is modified by a term proportional to $e^{4\si_0}$. Then, the modified Eq.~(\ref{loeomj}) sets the value of $\si_0$, rather than imposes a fine-tuned condition on the theory parameters.

It has been shown that the GR-like black holes in the dRGT model suffer from problems such as instabilities and strong coupling scales \cite{Berezhiani:2011mt,Berezhiani:2013dw,Berezhiani:2013dca,Kodama:2013rea}. It would be interesting to investigate the perturbations on the GR-like black holes in the quasi-dilaton models, which are the only static, spherical symmetric black holes in the theory.

It may well be the case that these GR-like black holes in (generalized) quasi-dilaton massive gravity also suffer from similar problems as in the dRGT model. However, it is worth emphasizing that this does not mean that there are no phenomenologically valid black hole solutions in (generalized) quasi-dilaton massive gravity. Indeed, even in the dRGT model, the natural black holes may pick up some mild time dependence and/or slightly deviate from spherical symmetry, the effects of which are suppressed by the small graviton mass. Those solutions are much more difficult to handle either analytically or numerically. Some interesting (partial) solutions of this class are the discharging black holes in the dRGT model discussed in \cite{Mirbabayi:2013sva}. These mildly non-static black holes are around the $\Lambda_2$ backgrounds of massive gravity \cite{deRham:2015ijs,deRham:2016plk}, which are actually closer to the original idea of the Vainshtein mechanism. It is also expected that the situation is similar for inhomogeneous/anisotropic cosmological solutions in massive gravity.

\vskip 20pt

\textbf{Acknowledgments}
We would like to thank Yun-Song Piao and Andrew J.~Tolley for discussions. SYZ acknowledges support from DOE grant DE-SC0010600. DJW is supported by NSFC, No.~11222546, and National Basic Research Program of China, No.~2010CB832804, and the Strategic Priority Research Program of Chinese Academy of Sciences, No.~XDA04000000.

\end{document}